\documentclass[fleqn,usenatbib]{mnras}

\usepackage{graphicx}	
\usepackage{amsmath}	
\usepackage{amssymb}	
\usepackage{multicol}        
\usepackage{bm}		
\usepackage{pdflscape}	
\usepackage{mathptmx}

\usepackage[T1]{fontenc}
\usepackage{ae,aecompl}

\usepackage{newtxtext,newtxmath}

\newcommand\hst{{\it HST}}
\newcommand\chandra{{\it Chandra}}
\newcommand\sax{{\it BeppoSAX}}
\newcommand\suzaku{{\it Suzaku}}
\newcommand\xmm{{\it XMM-Newton}}
\newcommand\asca{{\it ASCA}}
\newcommand\rxte{{\it RXTE}}
\newcommand\astrosat{{\it AstroSat}}
\newcommand\nustar{{\it NuSTAR}}
\newcommand\einstein{{\it Einstein}}
\newcommand\exosat{{\it EXOSAT}}
\newcommand\ginga{{\it Ginga}}

\newcommand\kev{{\rm~keV}}
\newcommand\keV{{\rm~keV}}

\newcommand\msun{{\rm~M_{\odot}}}

\title[\astrosat{}/UVIT observations of IC~4329A]{\astrosat{}/UVIT observations of IC~4329A: Constraining the accretion disc inner radius}

\author[G. C. Dewangan et al.]{
Gulab C. Dewangan,$^{1}$\thanks{E-mail:gulabd@iucaa.in}
P. Tripathi,$^{1}$ 
I. E. Papadakis,$^{2,3}$
and K. P. Singh$^{4}$
\\
$^{1}$Inter-University Centre for Astronomy and Astrophysics (IUCAA), SPPU Campus, Pune, India \\
$^{2}$Department of Physics and Institute of Theoretical and Computational Physics, University of Crete, 71003 Heraklion, Greece\\
$^{3}$ Institute of Astrophysics - FORTH, N. Plastira 100, 70013 Vassilika Vouton, Greece \\
$^{4}$Indian Institute of Science Education and Research Mohali, Knowledge City, Sector 81, Manauli P.O., SAS Nagar, 140306, Punjab, India
}

\date{Accepted XXX. Received YYY; in original form ZZZ}

\pubyear{2021}

\begin{document}
\label{firstpage}
\pagerange{\pageref{firstpage}--\pageref{lastpage}}
\maketitle

\begin{abstract}
We present a study of far and near ultraviolet emission from the accretion
disk in a powerful Seyfert 1 galaxy IC~4329A using observations performed with the Ultraviolet Imaging Telescope (UVIT) onboard \astrosat{}. These data provide the highest spatial resolution and deepest images of IC~4329A in the far and near UV bands acquired to date. The excellent spatial resolution of the UVIT data has allowed us to accurately separate the extended emission from the host galaxy and the AGN emission in the far and near UV bands. We derive the intrinsic AGN flux after correcting for the Galactic and internal reddening, as well as for the contribution of emission lines from the broad and narrow line regions. The intrinsic UV continuum emission shows a marked deficit compared  to that expected from  the ``standard'' models of the accretion disk around an estimated black hole mass of $1-2\times10^8M_{\odot}$ when the disk extends to the innermost stable circular orbit. We find that the intrinsic UV continuum is fully consistent with the standard disk models, but only if the disk emits from distances larger than  $\sim 80-150$ gravitational radii.
\end{abstract}

\begin{keywords}
galaxies: active -- galaxies: individual (IC~4329A) -- galaxies: Seyfert 
\end{keywords}

\section{Introduction} 
\label{sec:intro}

The main ingredient of the accretion process responsible for the extraordinary
luminosity of active galactic nuclei (AGN) is the presence of an accretion
disc around a supermassive black hole (BH) at the center of an active galaxy.
Observationally, the accretion disc emission is believed to be responsible for the Big Blue Bump (BBB)
continuum  spectral component \citep{1983ApJ...268..582M,1987ApJ...321..305C,1994ApJS...95....1E,1999PASP..111....1K,2005ApJ...619...41S}. This component extends from near-infrared at $\sim 1\mu$ to far and extreme ultraviolet at $\sim$1000\AA, and accounts for more than half of the bolometric luminosity in type 1 AGN. However, the nature of the accretion discs in AGN is not as clear as in the case of black hole X-ray binaries where the optically thick and geometrically thin standard accretion disc models have been successful
\citep{2006csxs.book..157M,Davis_2006}.  A great deal of complexity in AGN arises due to a number of effects such as contamination from host galaxy and numerous emission lines from the broad and narrow-line regions, intrinsic extinction, lack of high quality ultraviolet data, etc.   

Observations of relativistically broadened iron K$\alpha$ line and the related X-ray reflection emission from a number of AGN have indirectly revealed the presence of optically thick material in  the form of discs around supermassive blackholes \citep[e.g.,][]{2000PASP..112.1145F,2007ARA&A..45..441M}. These observations, however, cannot probe the radial dependence of the temperature and emission spectrum that are essential to unravel the nature of the accretion discs in AGN. Thus, the study of the intrinsic ultraviolet emission is crucial to probe the accretion discs in AGN. The Ultra-Violet Imaging Telescope \citep[UVIT;][]{2017AJ....154..128T,2020AJ....159..158T} on-board the \astrosat{} mission \citep{2014SPIE.9144E..1SS} with far and near UV sensitivity and excellent spatial resolution (full width at half maximum, FWHM$\sim1-1.5{\rm~arcsec}$), provides a unique opportunity to probe the intrinsic accretion disc emission from nearby AGN.  

We studied the UV emission from the IC~4329A active nucleus using \astrosat{}/UVIT observations. IC~4329A is a nearby Seyfert 1.2 AGN \citep{2006A&A...455..773V} at a redshift of $z = 0.016054$ \citep{1991AJ....101...57W}. It is the 2nd brightest type 1 AGN after NGC~4151 in the Swift/BAT catalog \citep{2013ApJS..207...19B}. The AGN resides in an edge-on galaxy with a dust lane passing through the nucleus. Based on a detailed multi-wavelength spectral study 
\citet{2018A&A...619A..20M} estimated an accretion rate of $10-20\%$ of the Eddington rate given its black hole mass in the range of $1-2\times10^8\msun$. IC~4329A is extensively studied in the X-ray band by all major satellites e.g., \einstein{}~\citep{1984ApJ...280..499P, 1989ESASP.296.1105H}, \exosat{}~\citep{1991ApJ...377..417S}, \ginga{}~\citep{1990ApJ...360L..35P},  \asca{}/\rxte{}~\citep{2000ApJ...536..213D}, \sax{}~\citep{2002A&A...389..802P}, \chandra{}~\citep{2004ApJ...608..157M}, \xmm{}~\citep{2007MNRAS.382..194N}, \suzaku{}~\citep{2016MNRAS.458.4198M}, and most recently with \nustar{}~\citep{2014ApJ...788...61B}.  

We describe our \astrosat{} observations and data reduction in Section~\ref{sec:obs}, followed by imaging analysis and derivation of intrinsic AGN emission in Section~\ref{sec:ana}. We derive the accretion disk spectra in section~\ref{sec:uv-optspec} and discuss our results in Section~\ref{sec:discuss}.

\begin{table}
  \centering
  \caption{AstroSat/UVIT observations of IC~4329A}
    \label{uvit_obs_list}
    \begin{tabular}{cccc}
      \hline\hline
       Obs. ID & Date   & \multicolumn{2}{c}{Exposure time (ks)} \\  
               &           & FUV/F154W &NUV/N245M \\ 
      \hline
  9000001006 &2017-02-03 &18.8&17.2\\  
  9000001048 &2017-02-23  &18.1 &20.0\\  
  9000001118 &2017-03-29  &19.5&19.6 \\  
  9000001286&2017-06-11  &18.2 &18.8\\  
  9000001340&2017-06-25  &12.4 &18.4\\  
      \hline\hline    
    \end{tabular}
  \label{tab:obs}
\end{table}


\begin{figure*}
\centering
\includegraphics[scale=0.75]{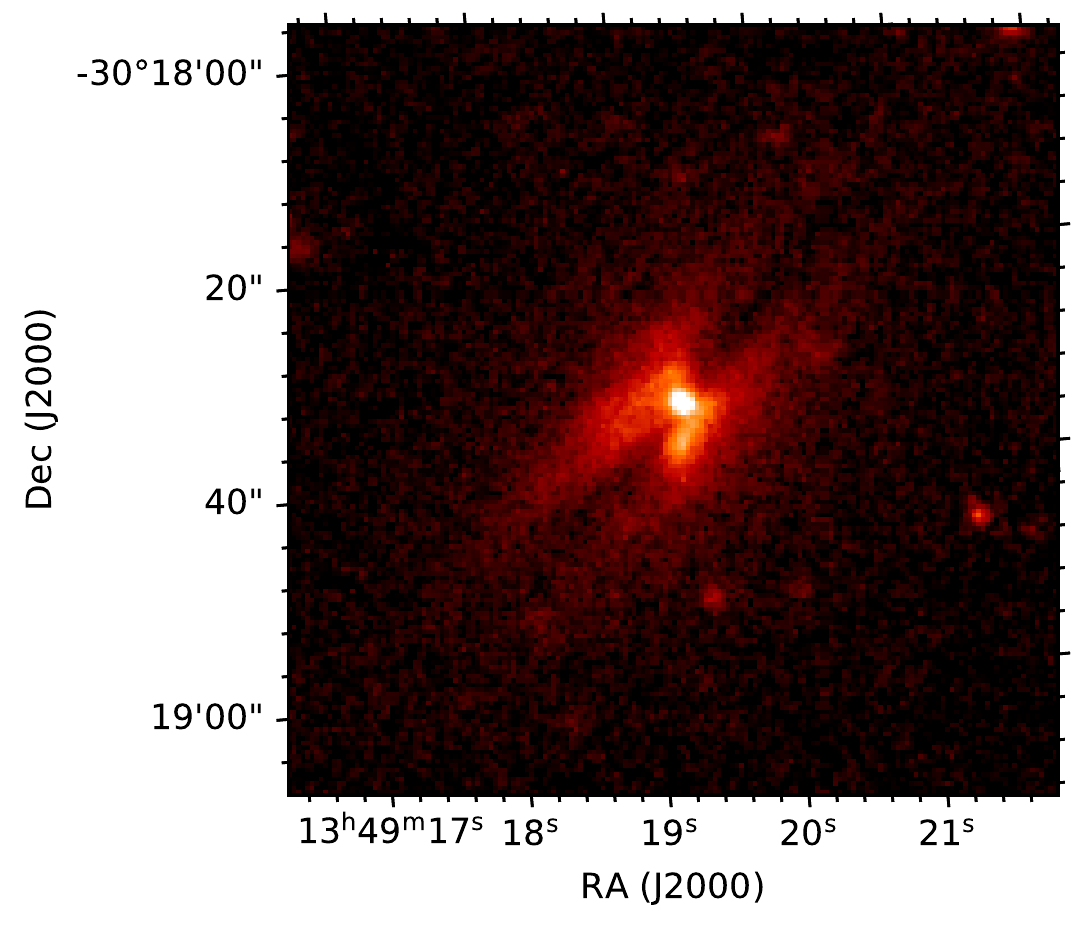}
\includegraphics[scale=0.8]{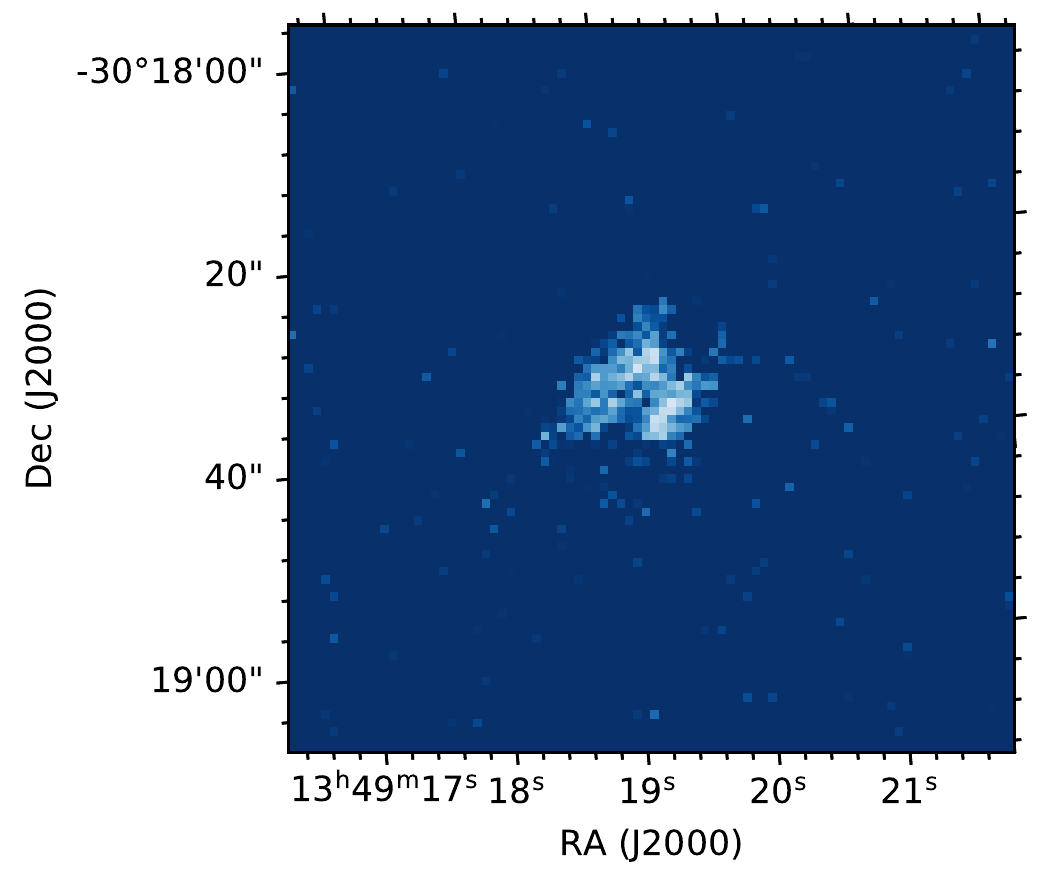}
\includegraphics[scale=0.5]{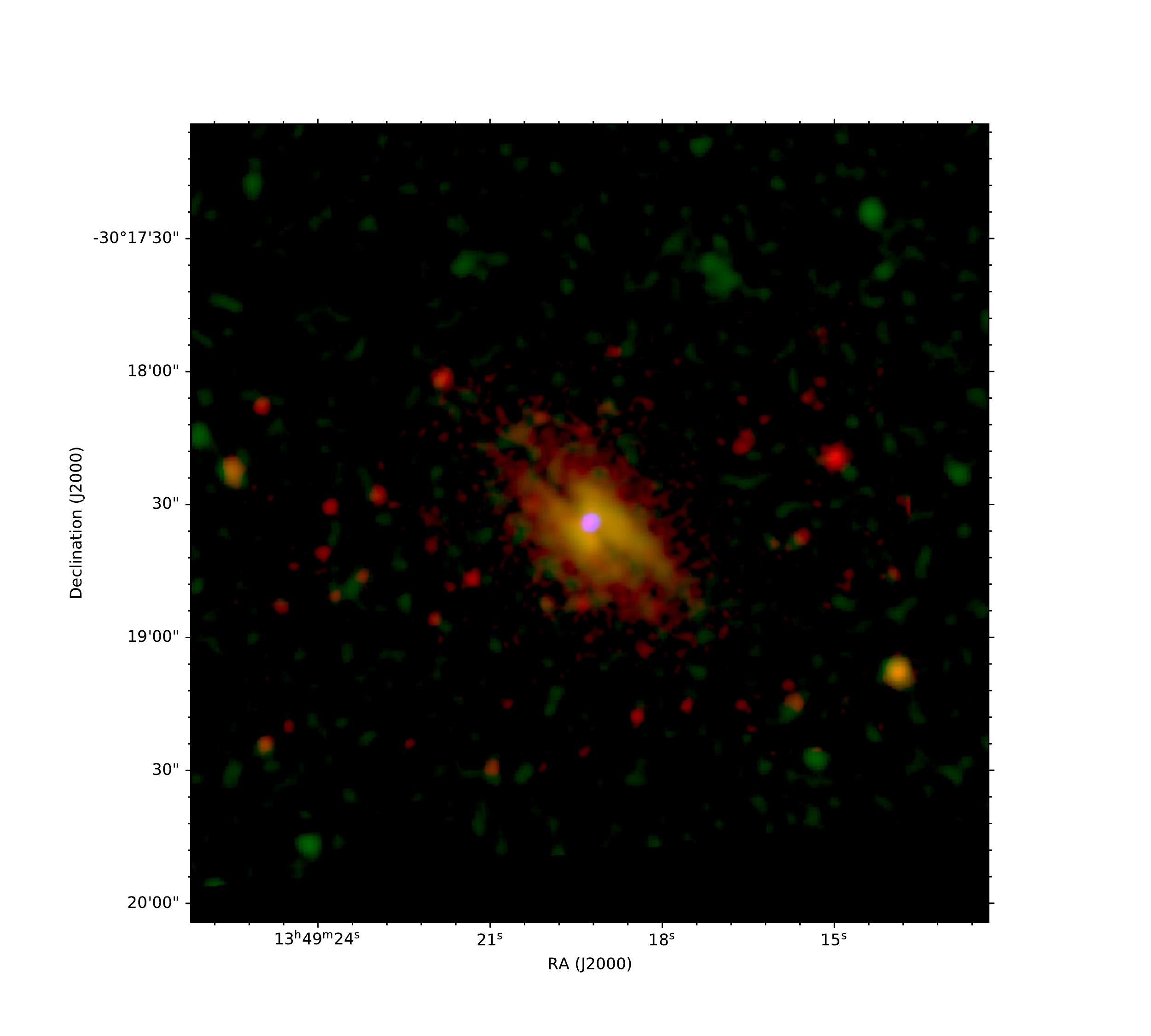}
\caption{{\it Upper panels:} The NUV and FUV \astrosat{}/UVIT images of IC~4329A (left and right panel, respectively). The unresolved active nucleus is clearly seen in the NUV but not in the FUV band due to heavy extinction by the dust lane in the host galaxy passing through the central regions.
{\it Bottom panel:} Composite  \astrosat{}/UVIT NUV (red), FUV (green) and \chandra{} X-ray (blue) image of IC~4329A. The nucleus is clearly detected in X-rays as well.} 
\label{fig:color_image}
\end{figure*}

\section{Observations \&  Data Reduction} 
\label{sec:obs}

IC~4329A was observed with \astrosat{} five times during February to June 2017
(Table~\ref{tab:obs}). Here we present the data from the Ultra-Violet
Imaging Telescope (UVIT) which was operated in the Photon-Counting mode using
two broadband filters: FUV/BaF2 (F154W, $\lambda_{mean}$ = 1541\AA{}, $\Delta \lambda$ = 380\AA{}) and NUV/NUVB13 (N245M, $\lambda_{mean}$ =2447\AA{}, $\Delta \lambda$ = 280\AA{}). We obtained the {\tt level1} data 
from the \astrosat{} data archive\footnote{\url{https://astrobrowse.issdc.gov.in/astro_archive/archive/Home.jsp}}, and processed them using the UVIT pipeline CCDLAB \citep{2017PASP..129k5002P}. We generated cleaned images for each observation and merged the five images for each filter to increase the signal-to-noise ratio (SNR). The net exposure time of the resulting image is 94~ks (NUV) and 87~ks (FUV). 
We derived the astrometric solution transforming the image coordinates to the world coordinates using the {\tt astrometry.net} package \citep{2010AJ....139.1782L}. 

We show the NUV and FUV images in Figure~\ref{fig:color_image} (upper panels). The edge-on host galaxy is seen in both images. A dust lane also appears, passing through the central nucleus. This is also observed with HST \citep{2018A&A...619A..20M}. We did not detect the active nucleus in the FUV image. In order to find its position in this image, we generated a composite three-color image  using the FUV and NUV data, as well as X--ray data from {\it Chandra}. We obtained the publicly available \chandra{}/HETG data acquired from the observation performed on 2017 June 6 (observation ID:20070) and generated the zero order X-ray image from the {\tt level2} event file. The composite color image is shown in Fig.~\ref{fig:color_image} (lower panel). The position of the bright nucleus in the X-ray and near UV band indicates that the dust lane in the host galaxy passes over the nucleus, and obscures its emission in the far UV band. Based on the astrometry and the composite three color image, we identify the location of the AGN in the FUV image which we use to generate the radial profile below.

\section{Data Analysis} 
\label{sec:ana}
The observed UV/optical emission from active galaxies not only consists of the intrinsic accretion disc emission but it also includes emission from the host galaxy, the broad and narrow-line region, the Fe~II emission complex and the Balmer continuum.  In addition, the observed flux is also affected by the internal and the Galactic extinction. Below we describe the steps we followed to derive the intrinsic continuum emission.

\subsection{Host galaxy Contribution}  

\begin{table*}
\centering
\caption{Results from the radial profile analysis of IC~4329A}
\begin{tabular}{llllll}  \hline
 \multicolumn{2}{c}{Model  Parameter} & \multicolumn{2}{c}{NUV} & \multicolumn{2}{c}{FUV} \\ 
             &        & Star & IC~4329A  &  Star & IC~4329A \\ \hline
    Moffat & $\alpha$  & $1.81\pm0.05$  & $1.81$(f) & $2.05\pm0.08$  & $2.05$(f) \\
           & $\gamma$  & $0.88\pm0.03$ &  $0.88$(f) & $1.18\pm0.05$   & $1.18$(f) \\
           & $A_m^a$ & $2792.7\pm79.7$  & $2461.3\pm41.1$ & $1621.0\pm48.7$ & $<32.5$ ($3\sigma$ upper-limit) \\
    Constant    & $c_0^a$  & $46.8\pm0.2$ & $46.8$(f)  &  $18.3\pm0.1$  & $18.3$(f) \\ 
     Exponential &   $\xi$ & -- & $-0.208\pm0.005$ & -- & $-0.19\pm0.01$ \\
        &   $A_e^a$ & -- & $462.8\pm13.8$ &-- & $61.0\pm8.5$ \\
     Gaussian    &  $x_0$  & -- &  -- & --&  $3.0\pm0.2$ \\
        & FWHM & -- &  -- &  -- & $4.2\pm0.5$ \\
        &  $A_G^a$ & -- &  -- & -- & $39.0\pm3.3$ \\
         & $\chi^2/dof$  & $104.6/96$  &  $169.1/195$ & $59.0/86$ &  $181.5/174$     \\ \hline
\end{tabular}\\
$^a$ In units of counts arcsec$^{-2}$.
\label{tab:radprof_par}
\end{table*}

 In the UV band, dust extinction and star forming regions in spiral galaxies result in highly uneven and structured surface brightness in the observed images. It is difficult to model such  structured 2D surface brightness profiles with smooth model profiles. On the other hand, averaging the galaxy emission over all azimuthal angle results in a reasonably smooth radial profile, which can be used to model the host  galaxy and AGN profiles reliably.  We separated the AGN emission from  the host galaxy emission by fitting the radial surface brightness profile of IC~4329A with a galactic and a point source profile.
We generated  the radial profile of a star (see below) and the galaxy by considering the total number of counts per unit area in a large number of annular regions centered at each source position, up to a distance of $40\arcsec$ (this is necessary in order to determine well the background level in the galaxy radial profile). We used {\tt Sherpa} \citep{2001SPIE.4477...76F} to fit the radial profiles, and we quote $1\sigma$ errors on the best-fit parameters, corresponding to $68\%$ confidence level. 

We first derived the radial profile of a point source. For the NUV band, we chose the star TYC~7279--861--1 ($\alpha_{2000}=13h49m38.6$, $\delta_{2000}=-30d18m04.1s$), which is located outside the central region of the two images, shown in Fig.~\ref{fig:color_image}. This star is $\sim4.2{\rm~arcmin}$ away from IC~4329A, it is well isolated and bright enough for the radial profile analysis. TYC~7279--861--1 is faint in the FUV band, therefore we chose star CD--29~10609 ($\alpha_{J2000}=13h48m57.8s$, $\delta_{J2000}=-30d22m04.7s$) in the FUV band,   which is $\sim 5.75{\rm~arcmin}$ away from IC~4329A. This star is bright enough in the FUV band for the radial profile analysis. We also derived radial profiles for other point-like sources, and we verified that they have similar radial profiles.
We derived the point spread function (PSF) of the instrument by  fitting the radial profile of the point sources with a circular Moffat function: $f(r) = A_M (1 + (\frac{r}{\gamma})^2)^{-\alpha}$, and a constant, $c_{\rm 0}$, for the background. We accept the FWHM of this function as a measure of the instrumental resolution (FWHM= $2\gamma\sqrt{2^{1/\alpha} - 1}$).  We used the best-fit parameters of the Moffat function (Table~\ref{tab:radprof_par}) and we calculated the FWHM of
the PSF to be $1.2{\rm~arcsec}$ for the NUV/NUVB13 and $1.5{\rm~arcsec}$ for the
FUV/BaF2 filters. These values are similar to that reported by
\citet{2020AJ....159..158T,2017AJ....154..128T}. UVIT has superior
spatial resolution compared to Swift/UVOT  ($\sim
2.5{\rm~arcsec}$ FWHM; \citealt{2010MNRAS.406.1687B}) and \xmm{}/OM ($\sim2{\rm~arcsec}$ FWHM;  \citealt{2012MNRAS.426..903P})
in the UV band. 

We used the best-fitting Moffat function with variable nomalization to fit
the nuclear emission from IC~4329A. We also used an exponential profile, $I(r)
= A_E \exp{(\xi r)}$, and an off-centered Gaussian profile, $I(r) = A_G
\exp{(-\frac{4\log(2)(r - r_0)^2}{FWHM^2})}$, to model the host galaxy radial
profile. As before, we used a constant for the background.

The Gaussian profile is necessary to account for the off-centered
peak emission that is clearly observed in the FUV image. This could be due to emission by an additional, off-centered star forming region but, most probably, the off-centered peak in the radial profile appears due to the obscuration by the central dust lane, and the rapid decline in the host galaxy emission away from the center. If there was no obscuration, the galaxy profile would probably be well fitted by a single, exponential component. Now, with the obscuration due to the central dust lane, the off-centered Gaussian is necessary to account for the effects of the  obscuration in the central part\footnote{We note that our objective is not the study of the intrinsic properties of the host galaxy, therefore we do not fit the radial profile with an absorbed exponential profile but we separate the absorbed AGN emission from the host galaxy.}. In the NUV band,  the host galaxy is much brighter, due to significantly less obscuration and its stronger intrinsic flux. Therefore, an exponential gives a statistically good fit to the observed radial profile.

\begin{figure*}
\centering
  \includegraphics[scale=0.5]{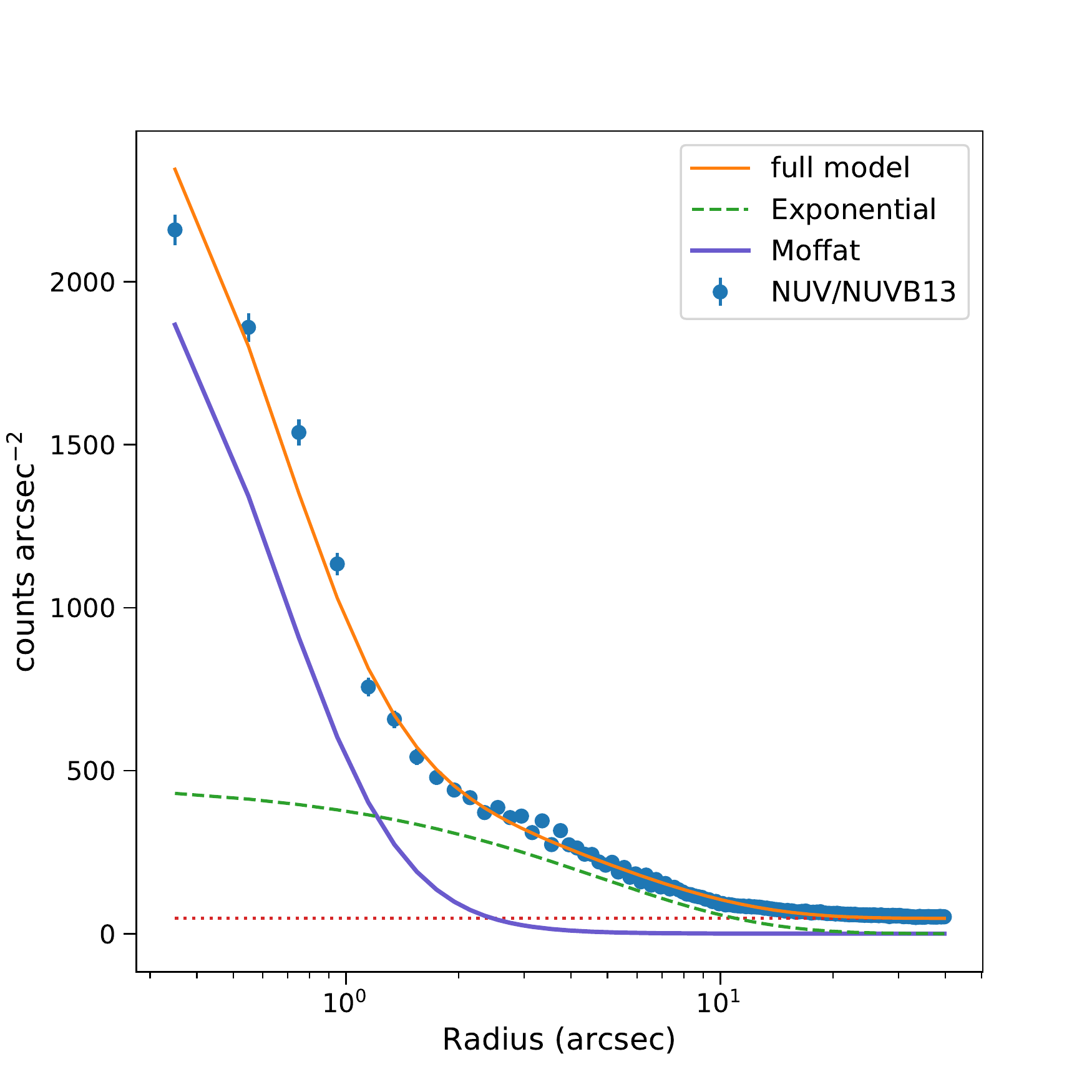}
  \includegraphics[scale=0.5]{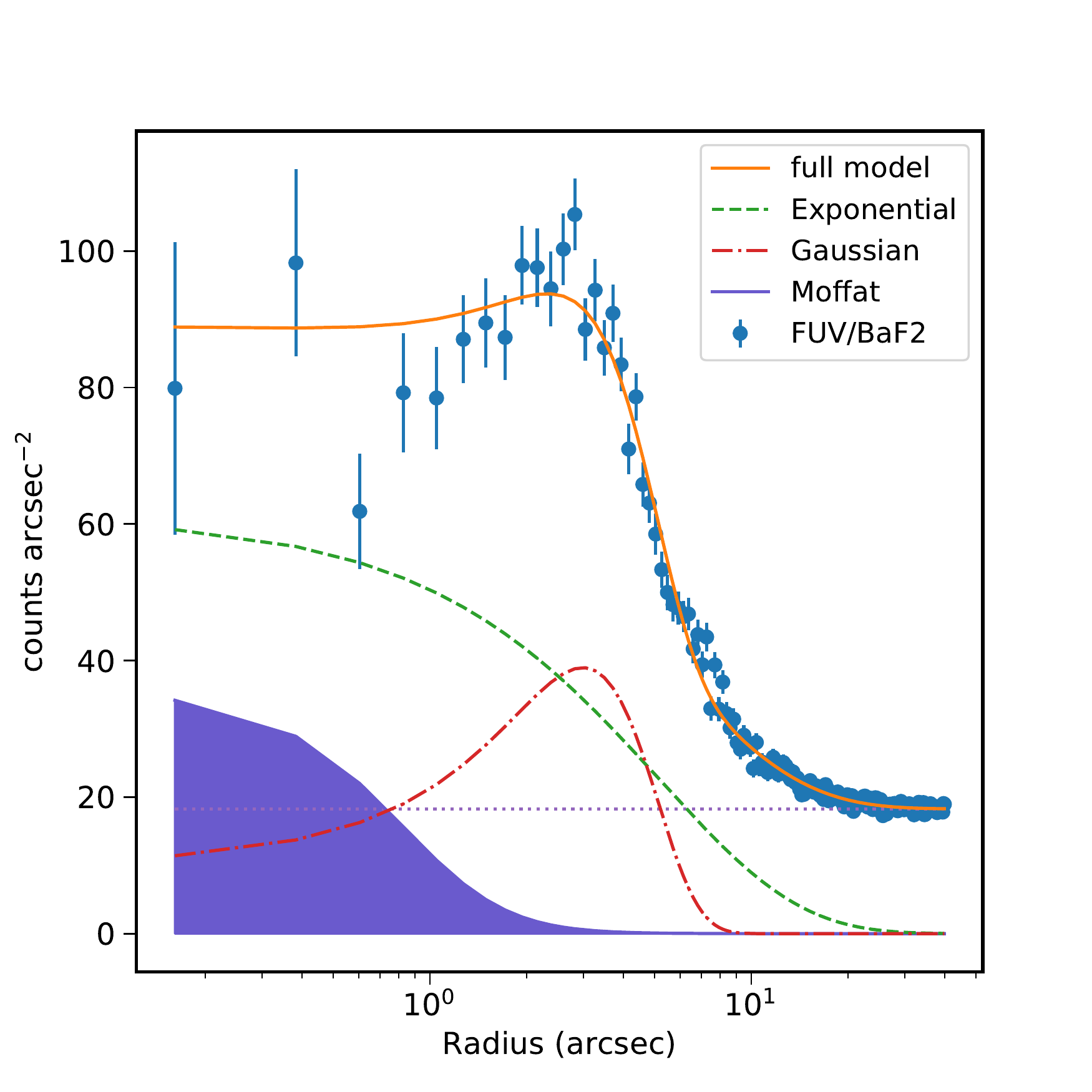}
  \caption{Radial profile of IC~4329A derived from the merged \astrosat{}/NUV ({\it left}) and FUV ({\it right}) images. The radial profiles are fitted with the instrument PSF for the AGN, and an exponential function for the galaxy emission plus a Gaussian function to account for the extra emission detected in the FUV band, probably due to the central star forming region. The AGN is not detected in the FUV band, and the shaded region in the right panel shows the $3\sigma$ upper limit on the AGN contribution. The deficit in the FUV emission in the central region is due to the the dust lane passing through the centre of the galaxy.}
  \label{fig:radprof}
\end{figure*}

The set of model components listed in Table~\ref{tab:radprof_par} resulted in a statistically good fit to the radial profile of IC~4329A in both the FUV and NUV bands.  The parameters of the best-fitting models are listed in Table~\ref{tab:radprof_par}. The radial profiles and the best-fitting models are shown in Figure~\ref{fig:radprof}. The best-fit Moffat amplitude is $0\pm10.8{\rm~couns~arcsec^{-2}}$ in the FUV band. This result shows that the AGN is not detected in this band (see also the upper right panel in Fig.~\ref{fig:color_image}). We list the $3\sigma$ upper limit on the Moffat amplitude in the last column of Table~\ref{tab:radprof_par}. 

We calculated the AGN count rate in each band by integrating the best-fit Moffat function within a
25 arcsec radius. We consider the 3-sigma upper limit on the Moffat normalization when calculating the FUV count rate. We converted the count rates to monochromatic flux densities using the unit conversion factors
derived from the calibration \citep{2017AJ....154..128T,2020AJ....159..158T}. 
The observed count rates and flux are listed in Table~\ref{tab:uvit_flux}. 

\subsection{Corrections for the Galactic and Internal Extinction} 

A significant amount of flux in the UV/Optical bands may be absorbed and
scattered by the interstellar dust grains both in the host galaxy as well as
in the ISM of the Milky Way. We corrected for the Galactic extinction using
the extinction curve of \citet{1989ApJ...345..245C}. We used the
line of sight color excess $E(B-V)=0.052$ from \citet{2011ApJ...737..103S} and fixed the ratio of total to selective extinction $R_V=\frac{A_V}{E(B-V)}=3.1$. As is obvious from the central dust
lane, the emission from the AGN in IC~4329A suffers significant internal
extinction. We describe below how we computed the unobscured intrinsic flux of the active nucleus. 

First, we corrected for the internal reddening using the extinction
curve of \citet{2004MNRAS.348L..54C} that lacks the $2175$~\AA{}
bump and is flatter than that of the Milky Way. This extinction curve was
constructed  based on the blue and red composite quasar spectra, and is found
to be appropriate for reddened AGN \citep{2012A&A...542A..30M,2018A&A...619A..20M} including IC~4329A.  

The internal reddening can be determined based on the Balmer decrement i.e., the ratio of the observed flux of the Balmer lines H$\alpha$ to H$\beta$.  Three measurements \citep{1992ApJ...393..658M,1996ApJ...469..648W,2017ApJ...846..102M}
in the last 30~years show that the  Balmer decrement in IC~4329A is in the range of $8.85-9.5$. For $R_V=3.1$, the Czerny et al. extinction law leads to the relation between reddening and the Balmer decrement, $E(B-V)=1.475\log{(R_{obs}/R_{int})}$, where $R_{obs}$ and $R_{int}$ are the observed and intrinsic H$\alpha$/H$\beta$ ratios. We used 
$R_{int}=2.72$, which is a value derived for observations of unreddened AGN \citep{2017MNRAS.467..226G}.  For $R_{obs}=8.85-9.5$, we found $E(B-V)=0.75-0.8$ for IC~4329A.  We adopt the internal reddening of $E(B-V)=0.8$ based on the largest value of Balmer decrement.  By fitting the broadband continuum from IR to near UV, \cite{2018A&A...619A..20M} measured an internal extinction of $E(B-V)=1.0\pm0.1$ for IC~4329A. Hence, we also used $E(B-V)=1.0$ and derived two sets of intrinsic flux for $E(B-V)=0.8$ and $1.0$. 

We also considered the extinction curve of \citet{2007arXiv0711.1013G} found to be appropriate for AGN \citep{2007ASPC..373..586C}. For this curve, the relation between the reddening and the Balmer decrement can be written as  $E(B-V)=1.79\log{(R_{obs}/R_{int})}$ for $R_V=3.1$. Using $R_{obs}=9.5$ and $R_{int}=2.72$, we derived $E(B-V)=0.97$ which we used to deredden the observed flux. The presence of dust lane and UV emission from  the host galaxy suggests strong star formation. Thus, the attenuation of AGN light in IC~4329A may be due to dusty environment in the central star-forming region, within the host galaxy. Therefore we also used the extinction of law of \citet{2000ApJ...533..682C} derived for local starburst galaxies. We found $E(B-V)=1.07$ for $R_{obs}=9.5$ and $R_{int}=2.72$, using the relation  $E(B-V)= 1.94\log{(R_{obs}/R_{int})}$ for $R_V=4.05$ derived from the \citet{2000ApJ...533..682C} extinction law. 


\begin{figure}
\includegraphics[scale=0.5]{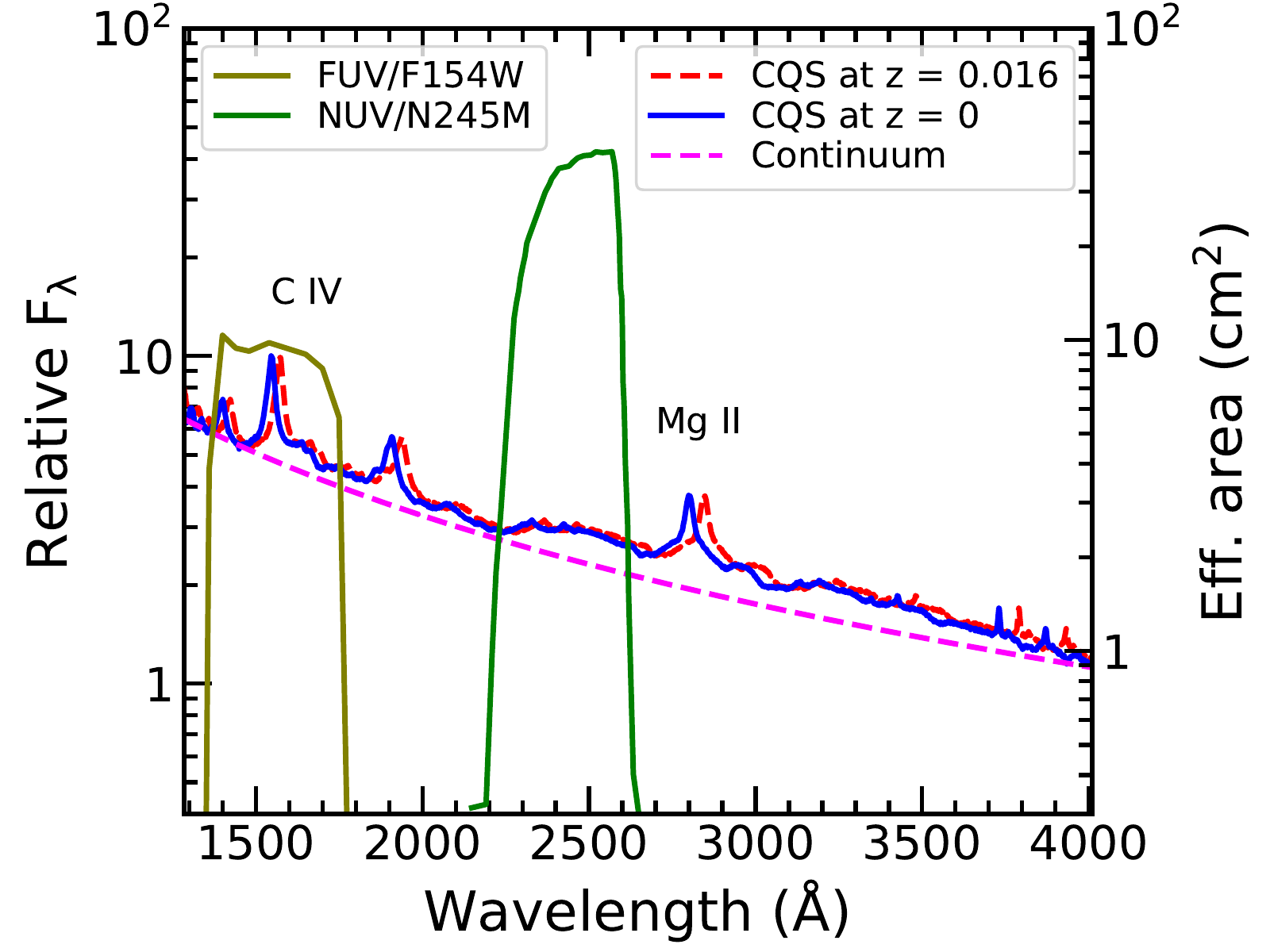}
\caption{
The composite quasar spectrum (CQS) of \citet{2001AJ....122..549V}, shifted to match the redshift of IC~4329A. Also shown are the effective areas of FUV/F154W and NUV/N245M \citep{2020AJ....159..158T}. The dashed-line show the continuum that was used to derive the contribution of emission lines in the two broadband filters.}
\label{cqs}
\end{figure}

\begin{table*}
\caption{Far and near UV flux measurements of IC~4329A}
\begin{tabular}{lllll} \hline
 Quantity & Extinction law & E(B-V) & NUV & FUV \\ 
          &  -- & -- &     & ($3\sigma$ upper limit) \\ \hline
 Net AGN counts~s$^{-1}$ & -- & -- & $0.078\pm0.001$ &   $<0.0015$ \\
 Observed $f_{\lambda}^a$ & -- &  -- & $(5.7\pm0.1)\times10^{-2}$ & $<5.6\times10^{-3}$ \\
 Intrinsic$^b$ cont. $f_{\lambda}^a$ & \cite{2004MNRAS.348L..54C} & (0.8,1.0) & ($7.0\pm1.1$,$31.4\pm0.5$) &  ($<6.3$, $<35.2$)\\
                                     & \cite{2007arXiv0711.1013G}  & $0.97$ &   $28.2\pm0.5$  & $<11.3$  \\
                                         & \cite{2000ApJ...533..682C}   & $1.07$   &   $173.5\pm3$  &  $<163.5$  \\  \hline
 \end{tabular} \\
$^a$ In units of $10^{-15}$ ergs~cm$^{-2}$~s$^{-1}$~\AA$^{-1}$. \\
$^b$ Flux derived after dereddening due to the Galactic and internal extinction, and after correcting for the contribution of emission lines.
\label{tab:uvit_flux}
\end{table*}

\subsection{Correction for Emission-line contribution} 

Emission lines from the narrow-line and the broad-line region, including the mFe~II emission and Balmer continuum, may contribute significantly to the flux measured with broadband filters in the optical/UV range. In our case, the broad emission lines due to Si~IV, C~IV and Fe~II contribute predominantly to emission in the FUV/F154W filter bandpass, 
while the Balmer continuum and Fe II emission contribute to the NUV/N245M bandpass. 

To subtract the contributions from these components, we used the composite quasar spectrum derived by \cite{2001AJ....122..549V}.  We fitted a powerlaw model to the line-free $1350-1365$ and $4200-4230$\AA{} bands in the composite quasar spectrum, and derived the intrinsic continuum. The composite quasar spectrum
shifted to the redshift of IC~4329A and the best-fitting double powerlaw
continuum are shown in Figure~\ref{cqs}. Also shown are the effective areas of the broadband filters FUV/F154W and NUV/N245M.

We subtracted the best-fit continuum and derived the pure emission line spectrum. We then calculated the expected count rates in the FUV/F154W and NUV/N245M filters for both the emission-line spectrum and the composite quasar spectrum using the effective areas for the two filters. We thus calculated that emission lines contribute $17.7\%$ and $17.4\%$ to the total intrinsic flux in the FUV/F154W and NUV/N245M bands, respectively. We used these contributions to correct for the
contribution of the emission lines to the intrinsic UV flux of IC~4329A. Thus derived the intrinsic continuum flux density in the two filters
(dereddened, for all three extinction laws we considered, and free of host galaxy and emission lines contribution), are 
listed in Table~\ref{tab:uvit_flux}.

\section{The UV/optical spectrum of IC~4329A} 
\label{sec:uv-optspec}

\begin{figure*}
\centering
  \includegraphics[scale=0.65]{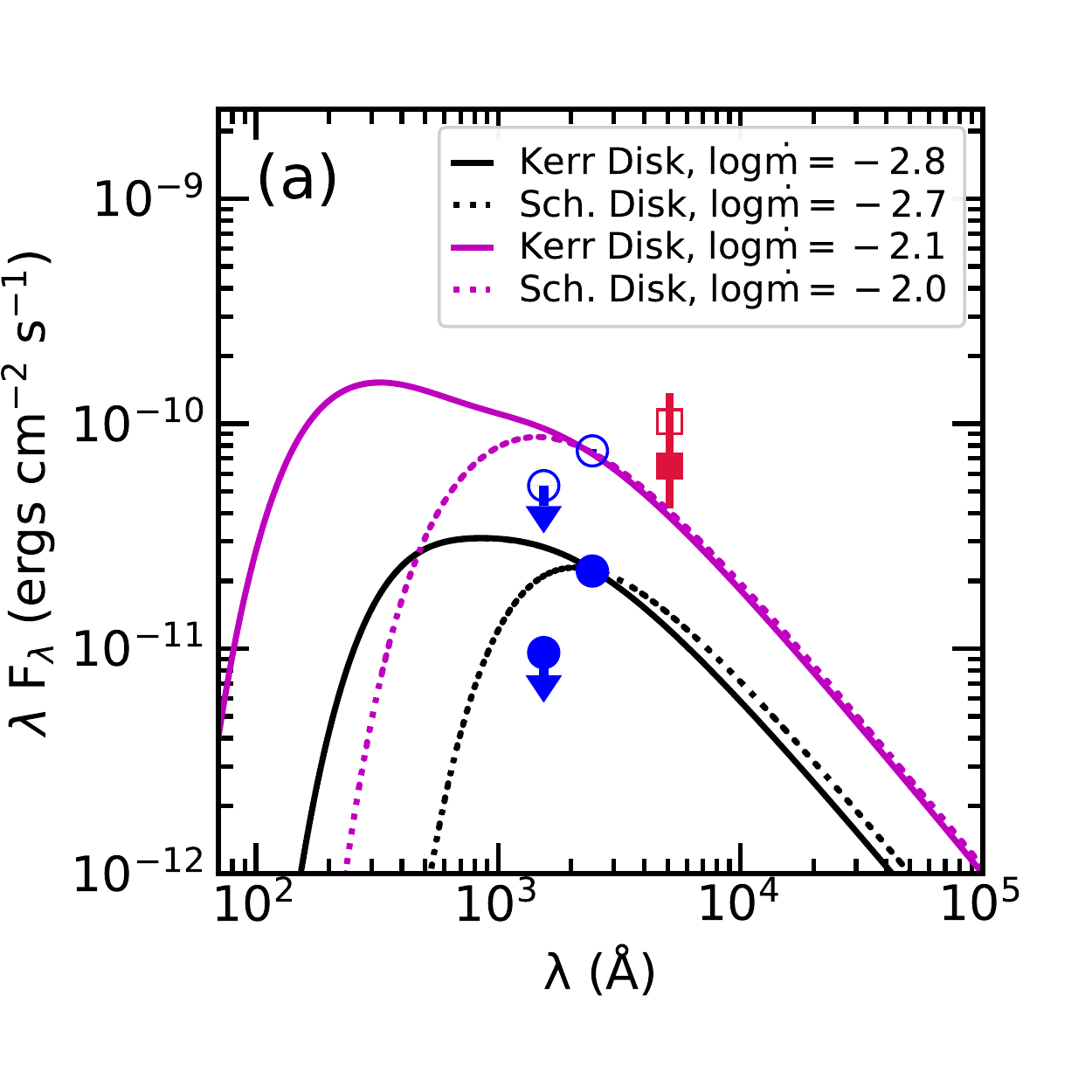}
  \includegraphics[scale=0.65]{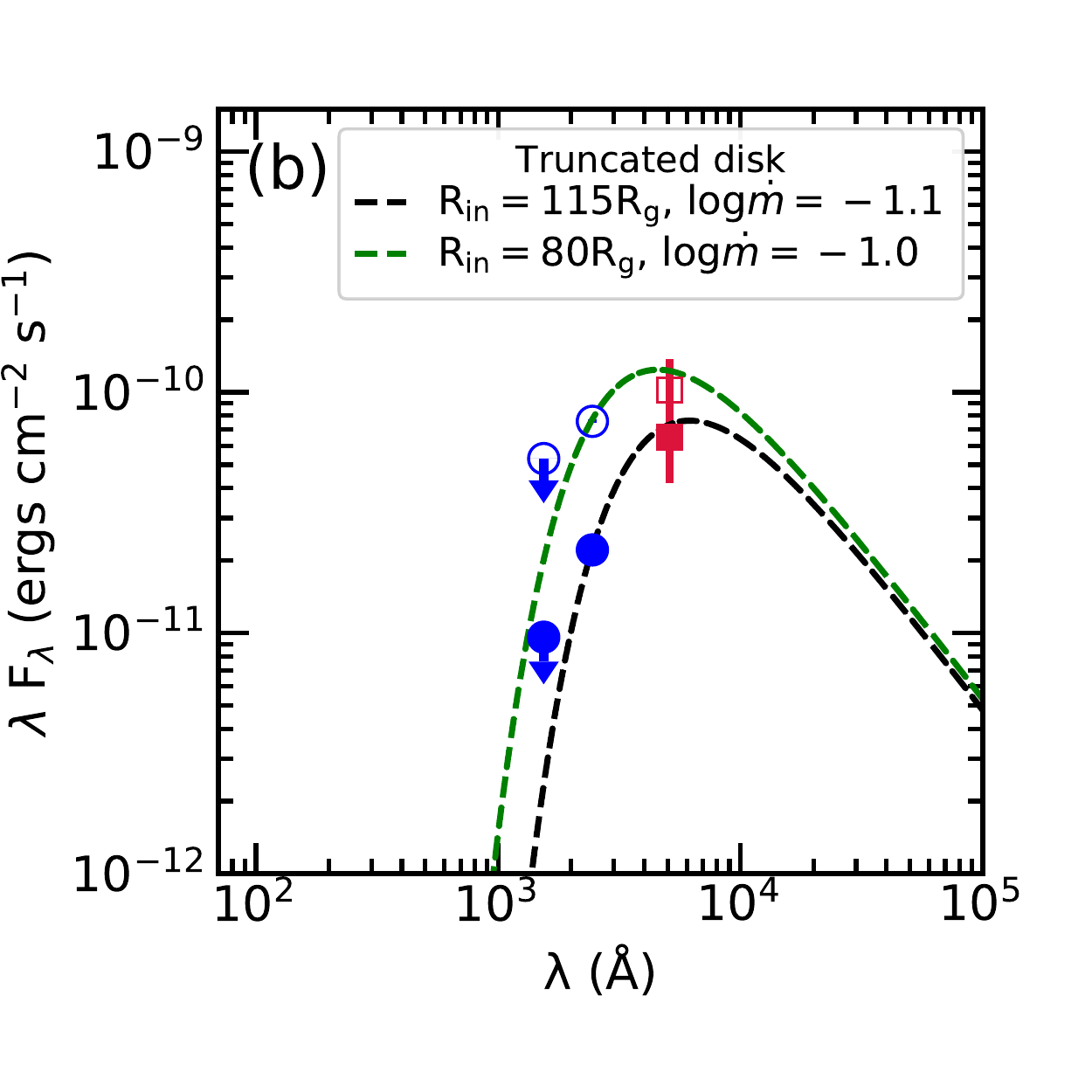}
  \includegraphics[scale=0.65]{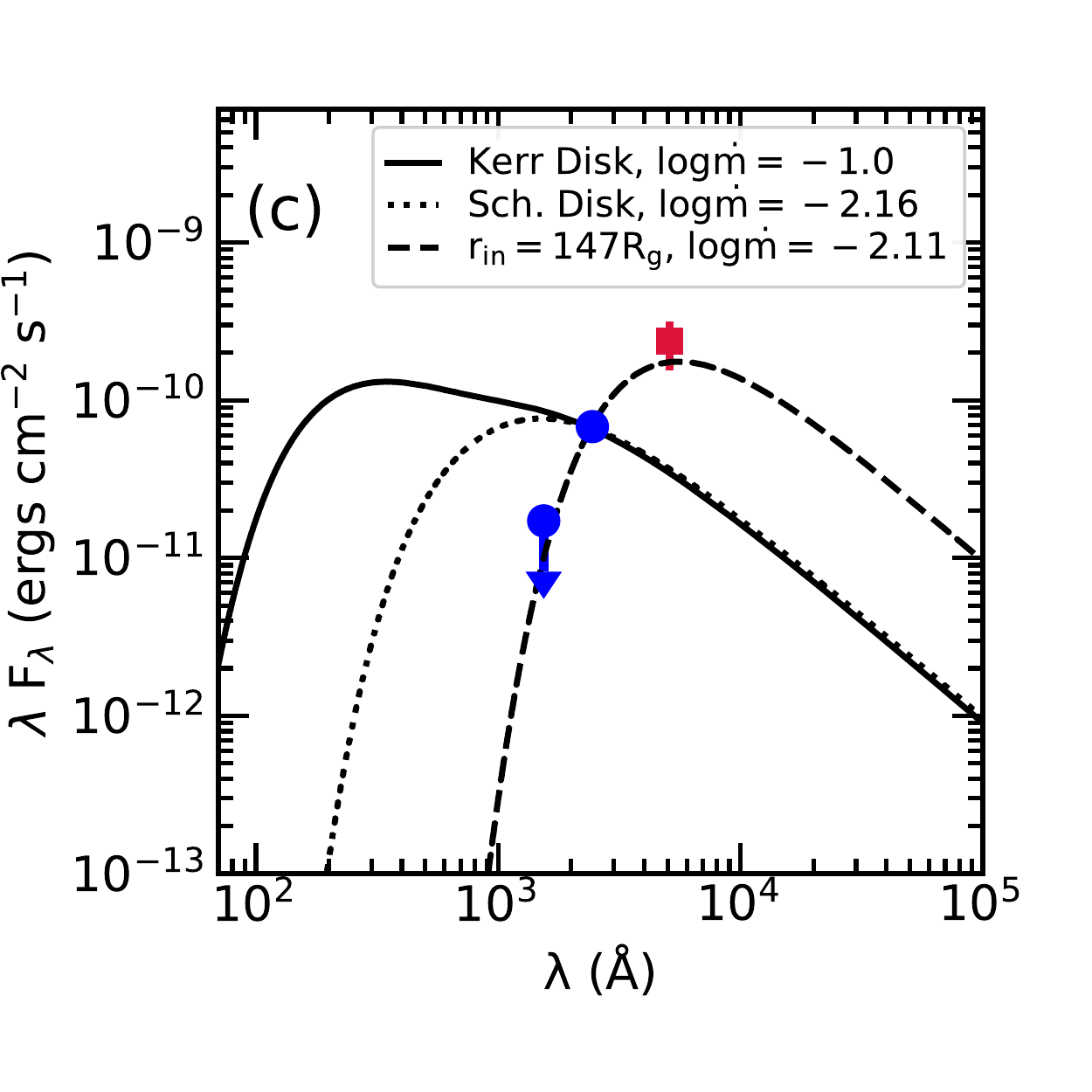}
  \includegraphics[scale=0.65]{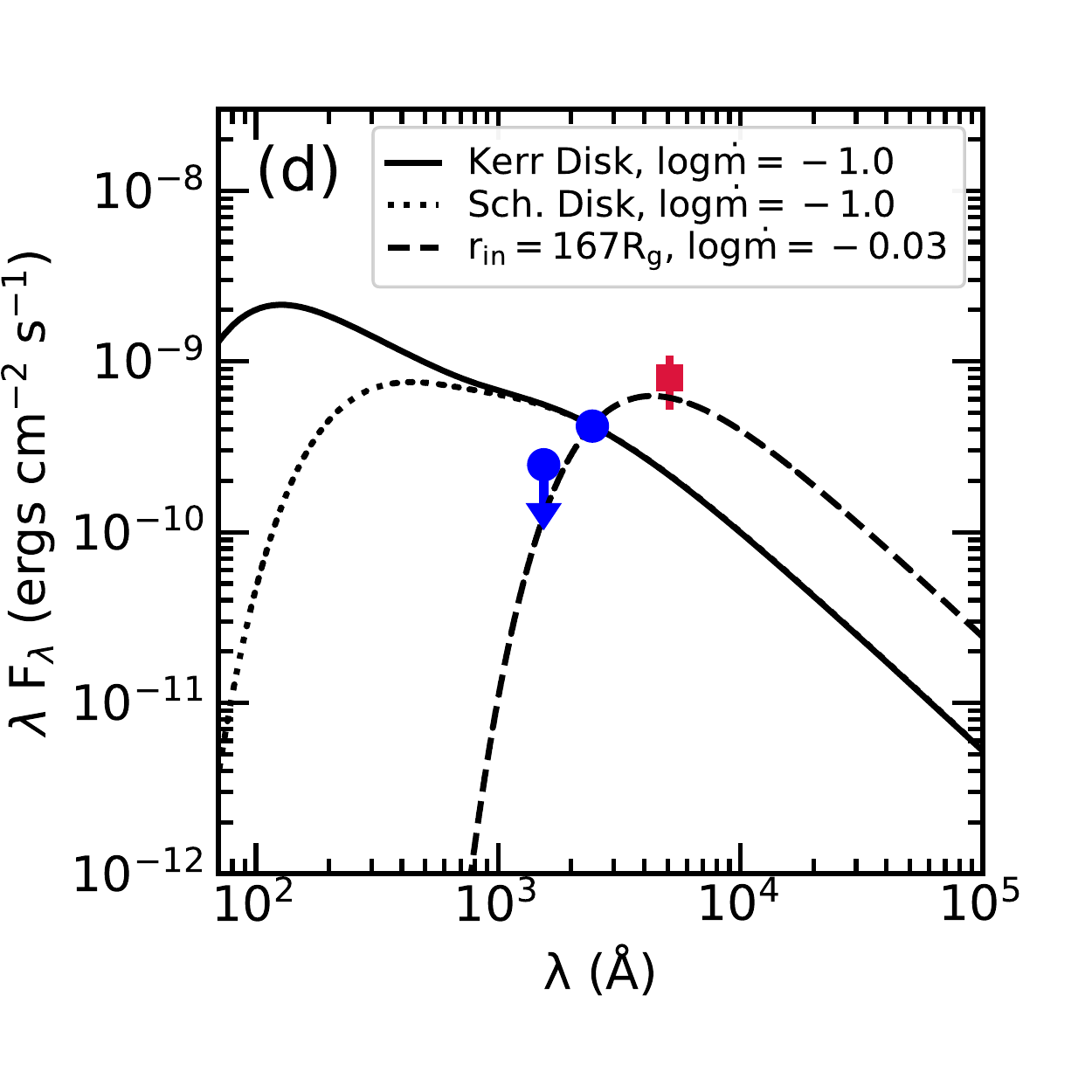}
\caption{The intrinsic, far and near UV flux measured with \astrosat{}/UVIT (circles) and the optical flux at 5100\AA{} based on the measurement of by \citet{2009ApJ...697..160B} (squares). The intrinsic AGN flux was derived using the extinction curves of \citet{2004MNRAS.348L..54C} (a,b),   \citet{2007arXiv0711.1013G} (c), and \citet{2000ApJ...533..682C} (d). 
In the case of the Czerny et al. extinction law, the intrinsic UV/optical fluxes were derived for $E(B-V)=0.8$ (filled symbols) and $E(B-V)=1.0$ (open symbols). Solid and dotted lines show theoretical accretion disk spectra when $R_{in}=R_{ISCO}$, in the case of a maximally rotating and a non-rotating, Schwarzschild black hole, respectively. The dashed lines show the theoretical spectra when the inner disc radius is larger than $R_{ISCO}$. Accretion rates (in Eddington limit units) and the inner disc radius listed in the inset box are those that give the best fit to the data (see text for details).}
  \label{fig:acc_disk_spec}
  \end{figure*}

We plot the intrinsic AGN flux in the far and near UV bands as a function of wavelength in all panels of Figure~\ref{fig:acc_disk_spec}. The results for the \cite{2004MNRAS.348L..54C}, \cite{2007arXiv0711.1013G} and the \cite{2000ApJ...533..682C} extinction curves  are plotted in the upper, lower left and lower right panels, respectively. Filled and open points in the upper panels show the flux measurements for the \cite{2004MNRAS.348L..54C} extinction curve derived for $E(B-V)=0.8$ and $1.0$, respectively. Squares in all panels of Figure~\ref{fig:acc_disk_spec} show the mean intrinsic flux at 5100\AA\ based on measurements by \cite{2009ApJ...697..160B}.  These
authors used a high resolution \hst{} image to decompose the galaxy contribution from the AGN  emission. They corrected the data for Galactic reddening, and applied all the additional corrections as we did to the UVIT data. They list the mean AGN flux using data from ground based, spectroscopic monitoring campaigns, thus taking into account the intrinsic variability of the source. In any case,  the optical variability amplitude of IC~4329A is of the order of few percent around the mean \citep{2018A&A...619A..20M}. Therefore, we can safely combine our and the \cite{2009ApJ...697..160B} measurements together, to construct the optical/UV spectral energy distribution (SED) of the source.

The black hole mass of IC~4329A has been estimated to be in the range of $\sim 1-2\times10^{8}\msun$ \citep{2009ApJ...698.1740M,2010A&A...524A..50D,2010MNRAS.402.1081V}. We assumed $M_{BH}=1.5\times10^{8}\msun$ and we computed spectra of
standard discs for accretion rates in the range between $\log{\dot{m}}=-3$  to
 $-1$, with a step of $\Delta \dot{m}=0.1$ (accretion rates are normalized to
the Eddington accretion rate). We used the  model {\tt OPTXAGNF} \citep{2012MNRAS.420.1848D} available in the XSPEC program (version 12.11.1; \citealt{1996ASPC..101...17A}  distributed with the heasoft package  (version 6.28), to calculate the disc spectra\footnote{We set the model parameter $r_{cor}$ equal to minus the inner disk radius to calculate only the disk spectrum.}. The model  assumes that the gravitational energy released in the disk at each radius is emitted as a (colour temperature corrected) blackbody using the Novikov-Thorne
prescription \citep{1973blho.conf..343N}. 

First we considered the case when the inner disc radius, $R_{in}$, is equal to $R_{ISCO}$, i.e. the radius of the innermost stable circular orbit (ISCO) for a maximally spinning Kerr black hole and a non-rotating Schwarzschild black hole ($R_{ISCO}=1.235 R_g$ and $R_{ISCO}=6R_g$, respectively; $R_g=GM_{BH}/c^2$ is the gravitational radius).
The solid and dashed lines in Figure~\ref {fig:acc_disk_spec} show the disc
spectra in the two cases. The model disc spectra shown are the ones with the
minimum difference (squared) with the data (including the upper limit in the
far-UV). The standard disc extending to ISCO fails to explain the observed data. As
expected, due to the $\nu^{1/3}$ dependency, the theoretical accretion disc
spectra rise towards shorter wavelength while the observed data show the
opposite trend. They show a steady flux decline, from the optical to the far-UV,
which is strongly emphasized by the upper limit in the FUV band.  

The discrepancy between the standard disc model and the data holds for all the different extinction laws we tried. Perhaps we can get a better agreement between the model and the observations if we assume a larger $E(B-V)$. Absorption by dust should affect the FUV band more than the NUV/optical bands. If we increase $E(B-V)$ the FUV upper limit will increase more than the NUV/optical flux measurements, and perhaps the data will be consistent with theory. We estimate $E(B-V)$ assuming $R_{int}=2.72$, which is a value consistent with theoretical calculations where the gas is optically thick in Lyman lines, in which case H$_\alpha/$H$_{\beta}\approx 2.85$ \citep{1938ApJ....88...52B, 2006agna.book.....O}. The intrinsic Balmer decrement can be different from one AGN to another, because it depends on the physical environment of the emission regions, e.g. hydrogen density. For example, Fig. 4 in \cite{2017MNRAS.467..226G} shows that $R_{int}\approx 1.9$ when log$N_{\rm H}\ge 12$. But, even in this case, we get $E(B-V)\sim 1$ if assuming the \cite{2004MNRAS.348L..54C} extinction curve for example, a case we have already considered in our calculations (empty points in the upper panels in Fig.~\ref{fig:acc_disk_spec}).

{\tt OPTXAGNF} is only an approximation of the disc emission in AGN. For one thing it does not include the relativistic effects in the inner disc region. These should be more pronounced in the case of maximally rotating BHs, where photons emitted from this region will suffer gravitational redshit, their trajectories will be strongly curved and may not reach the observer at infinity, etc. However, the discrepancy between model and data appear in a spectral region which should be relatively free of relativistic effects. The model spectra in the case of the maximally rotating and Schwarzschild BHs are identical below 1000\AA, for the BH mass we have assumed (see the solid and dotted lines in all panels of Fig.~\ref{fig:acc_disk_spec}). This implies that the emission at these wavelengths is dominated by radii larger than 6 $R_g$. In addition, the {\tt OPTXAGNF} model does not include the inclination effects, but these should mainly cause a change in the model normalization (due to the change of the projected disc area at high inclinations), at least at wavelengths longer than 1000\AA, where the relativistic effects should be less pronounced. However, we find a disagreement between the model and the observed slope, not between the predicted and the observed normalization. 

\begin{figure}
    \centering
    \includegraphics[scale=0.65]{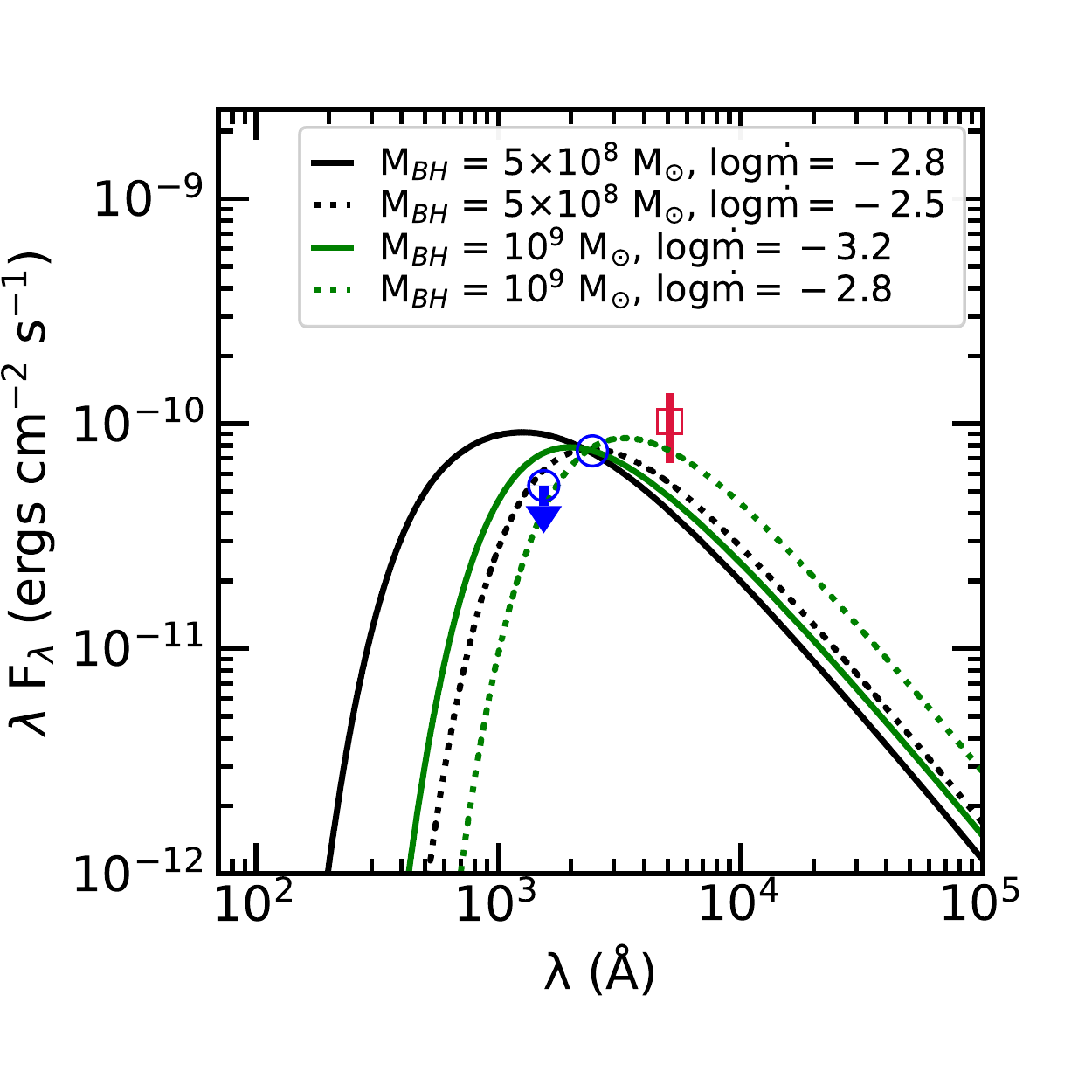}
    \caption{The Kerr (solid curves) and the Schwarzschild (dotted curves) accretion disks spectra for black-hole masses of $5\times10^{8}M_{\odot}$ (black) and $10^{9}M_{\odot}$ (green). AGN fluxes are corrected for the internal extinction using the extinction curve of \citet{2004MNRAS.348L..54C} with a color excess of $E(B-V)=1.0$.}
    \label{figBH_comparing_sed}
\end{figure}

Another possibility is that the BH mass is quite larger than the one we considered. We therefore derived Kerr and Schwarzschild disc spectra for masses equal to $5\times10^{8}M_{\odot}$ and $10^9M_{\odot}$, and for various mass accretion rates. The spectra that describe best the data are shown in Fig~\ref{figBH_comparing_sed}. Solid and dotted curves show the spectra for a maximally spinning and a non-rotating BH, respectively, while black and green colours show the spectra for the $M_{BH} = 5\times10^{8}M_{\odot}$ and $M_{BH} = 10^9M_{\odot}$. The figure shows that spectrum in the case of a Schwarzschild black-hole with a mass of $10^{9}M_{\odot}$ describes the data well (green dotted spectrum), with an accretion rate of $\log{\dot{m}} = -2.8$. However, this BH mass is almost ten times larger than all the other BH mass estimates for this galaxy (see for example the discussion on the BH mass estimation for this object in \cite{2018A&A...619A..20M}. 

Finally, the discrepancy between standard disc models and the data suggests that 
the possibility of a truncated accretion disc in IC~4329A. We
therefore generated theoretical spectra for accretion disks truncated at radii
larger than $R_{ISCO}$. We computed disk models with  with $R_{in}$ in the range
between 50 to 200 $R_g$, in steps of $\Delta R_{in}=1$, for the full range of
the accretion rates we considered. The dashed lines in the right panel of
Figure~\ref{fig:acc_disk_spec} show the best-fit disk spectra in this case, for a
Schwarzschild BH.  These are the model curves which resulted in the smallest
(squared) difference between model and data in the NUV and optical bands, and
are also consistent with the upper limit in the FUV band. Clearly, the data are
consistent with the spectrum emitted by truncated disks. The inner radius
depends on the assumed intrinsic extinction curve, and ranges from $R_{in}\sim
80$ in the case of the Czerny et al. curve, to $R_{in}\sim 150-170R_g$ when we
assume the other two curves. 

 \subsection{The inner disc}

Our results clearly demonstrate that the standard accretion disc models are consistent with the data but only if $R_{\rm in} \gtrsim 80-100 R_g$. The inner disc could be replaced by a hot, optically thin flow \citep[see,][]{2020MNRAS.491.5126M} 
  or there could exist a   warm, optically-thick medium on top of the inner disc,  as hypothesised in the thermal Comptonisation models for the origin of the soft X-ray excess \citep[see e.g.,][]{2012MNRAS.420.1848D}. In a companion paper, \cite{Tripathi_etal_2021} have studied the broadband ($0.3-20\kev$) X-ray spectra of IC~4329A acquired simultaneously with the \astrosat{} UV data presented here. They detected a soft-excess component, which can be fitted well either by a simple black-body component with $kT\sim 0.26\kev$, or by the thermal Comptonization model {\textsc{nthcomp}}, with a warm corona temperature of $kT_e\sim 0.26\kev$  and an optical depth of $\tau\sim27$. Their analysis also revealed that the fraction  of the continuum X-ray power-law  flux and the total X-ray flux (i.e. power-law plus soft-excess emission) to be $f_{PL}\sim 0.8$. 

We used the parameters for the soft excess and the power-law fraction and we refitted the  UV/optical data derived for the three different extinction curves with {\tt OPTXAGNF}. We kept the {\tt OPTXAGNF} parameters $kT_e$ and $\tau$, for the soft excess, and $f_{PL}$ fixed to the values derived by \cite{Tripathi_etal_2021}. We found that the three optical/UV SEDs are well described by the model when the black hole spin is fixed at $a=0$, while we could not fit the data in the case of $a=1$. The best-fit results are $R_{in}\sim 126$ and $\sim80$ for the  extinction curve of \cite{2004MNRAS.348L..54C} with $E(B-V)=1.0$ and $0.8$, and $R_{in}\sim200$, $74$  for the extinction curves of  \cite{2000ApJ...533..682C} and \cite{2007arXiv0711.1013G}, respectively. 

\cite{Tripathi_etal_2021} also measured the soft excess flux in the $0.3-2\keV$ band. They found it to be variable, between $1.1$ and $1.7 \times 10^{-10}$ ergs s$^{-1}$ cm$^{-2}$. We used the {\tt OPTXAGNF} best-fit models and we computed the model predicted soft-excess flux in the same energy range. We found that the best-fit soft-excess flux predictions are off by at least a factor of two from the values measured by \cite{Tripathi_etal_2021}, except for the extinction curve of \cite{2004MNRAS.348L..54C} with $E(B-V)=1.0$. In this case, the estimated soft excess flux in the $0.3-2\kev$ is within the range of the fluxes measured by \cite{Tripathi_etal_2021} from the five \astrosat{} observations. The best-fit inner radius and accretion rate are $126 R_g$ and  $\sim 2.4\%$ of the Eddington rate, respectively. This suggests that the inner disc in IC~4329A could be covered by a warm, optically thick layer, whose emission is consistent with the optical/UV SED and could also explain the observed soft-excess we detect in this source.



\section{Discussion and conclusions} 
\label{sec:discuss}

We present the results from the analysis of UV images of IC~4329A obtained with  \astrosat{}/UVIT observations. These are the highest resolution and deepest images of IC~4329A ever acquired with any telescope in the near and far UV bands.
The  UVIT images enabled us to calculate the galaxy radial profile, and unambiguously separate the AGN emission from the extended emission of the host galaxy by fitting the profiles with the appropriate functions. We derived the intrinsic AGN continuum flux after dereddening for both the Galactic and internal extinctions, and correcting for the contribution of emission lines.

The active nucleus is not detected in the FUV band, and we were able to put a conservative 3$\sigma$ upper limit on the AGN flux in this band. When combined with the AGN flux in the NUV and 5100\AA\, (taken from past observations), we found that the far-UV/optical, intrinsic emission of the active nucleus in IC~4329A, requires no disc emission below $\sim 80-170R_g$. The superb spatial resolution of
UVIT is crucial in this case, specially in the FUV. If we had performed
simple aperture photometry in the FUV image, using a 5-10~arcsec radius (as is
typically the case), we would have ended with a flux measurement at least 7
times larger than the $3\sigma$ upper limit we have determined in this band.
Since the host galaxy in most templates is assumed to emit very little at these
short wavelengths, this measurement would have been assumed to be intrinsic of
the active nucleus, and this false result would have critically altered the
shape of the intrinsic AGN spectral energy distribution (SED) at short
wavelengths. 

X-ray reflection spectroscopy can independently probe the inner accretion disks
via the  relativistic iron K$\alpha$ line  near $\sim 6.4\keV$ and the strength
of the reflection. Using \xmm{} observations of IC~4329A, \cite{2007MNRAS.382..194N} derived  a low disk reflection fraction of $R_f\sim 0.3$ ($R_f$ is defined as the ratio of the X--ray intensity that illuminates the disk to the intensity that reaches the observer; $R_f=1$ corresponds to a semi-infinite slab  subtending $2\pi$ solid angle at the X-ray source).  Using \nustar{} observations, \cite{2014ApJ...788...61B}  detected a moderately broad iron line with a Gaussian $\sigma \sim 0.36\kev$ in addition to narrow iron K$\alpha$, K$\beta$ lines. \cite{2019ApJ...875..115O} analyzed simultaneous
\suzaku{} and \nustar{} observations and showed that the iron line
complex is well described by relativistic reflection from an accretion disk
truncated at $R_{in}=87_{-13}^{+73}R_g$ and distant reflection from a clumpy
torus, with the relativistic reflection fraction $\sim 3\times 10^{-3}$. In general, the reflection fraction is unlikely to go below $\sim 0.5$ for an accretion disk extending to
the last stable orbit \citep{2016A&A...590A..76D}. Hence, the moderately broad
iron line and the low reflection fraction both support the hypothesis of a truncated disk in IC~4329A. 

The deficit of the intrinsic UV emission can arise if the disk is truly truncated i.e., there is no optically thick  material accreting in the inner regions to emit in the UV/optical bands. This is probably the case with Galactic, BH X--ray  binaries, when they are in their quiescent and hard-state (see \citet{1997ApJ...489..865E,2007A&ARv..15....1D,2020MNRAS.491.5126M}.
In this case, the inner disk is replaced by a hot, optically thin, ADAF--like
flow \citep{1994ApJ...428L..13N}, which emits in X--rays. This could also be the
case with low-luminosity AGN, like LINERS. However, this is supposed to happen
when the accretion rate is below $\sim 1-2$ percent of the Eddington limit.
Our truncated disk fits to the data indicate a higher accretion rate, of the
order of 0.1, or more, depending on the assumed extinction curve. 

Another possibility is that  the inner disk in IC~4329A is covered by a warm
layer ($kT\sim 0.2-0.5$ keV), with high optical depth of $\sim 10-20$. The idea of the warm, and optically thick corona, on top of the inner accretion disk, has been applied successfully to model the broad-band optical/UV/X--ray SEDs of several
AGN \citep{1998MNRAS.301..179M,2003A&A...412..317C,2012MNRAS.420.1825J,2015A&A...575A..22M,2018A&A...609A..42P,2018MNRAS.480.1247K,2020A&A...634A..85P}, although it is not certain yet whether such a layer/corona can  exist  in  the  upper  layers  of the inner accretion flow. If it does, and it is optically thick, all UV/optical photons emitted by the disk will be shifted to soft X--rays, resulting in a  deficit in the observed UV/optical emission.  We investigated this possibility by assuming the {\tt OPTXAGNF} model and using the soft excess parameters derived by \cite{Tripathi_etal_2021}. We found that the model predicted soft excess flux is consistent with the \cite{Tripathi_etal_2021} measurements only for one of the three extinction curves we considered. This is the \cite{2004MNRAS.348L..54C} extinction curve, with $E(B-V)=1$. This is in agreement with the results of \cite{2018A&A...619A..20M}. However, we infer an accretion ratio ratio of $\sim 2.4\%$ of the Eddington limit, while \cite{2018A&A...619A..20M} derived an accretion rate  of $\sim 10-20\%$, in Eddington units. Perhaps the higher Eddington fraction could be the reason why the optical-FUV spectral shape derived by\cite{2018A&A...619A..20M}  is relatively flat compared to our measurements.

The main result of our work is that the optical to far-UV SED of IC~4329A is
consistent with standard accretion disk models only if the inner disk does
not contribute to the observed UV emission.  The standard inner disk in IC~4329A
is either replaced by some optically thin, hot inner flow, or it is covered by an optically thick, "warm" X--ray corona. We cannot constrain  $R_{in}$ and the accretion rate accurately, mainly because we do not know the exact extinction curve. 

 If we assume that the observed soft excess \citep{Tripathi_etal_2021} is due to warm Comptonization, then we find that the extinction law is well described by \citet{2004MNRAS.348L..54C} with $E(B-V)=1$, the inner disc radius is $R\sim 125$ and the accretion rate is quite small ($\dot{m}\sim 0.025$).  Recently  \cite{2018MNRAS.480.3898N} have discovered that changing look AGN make spectral transitions at a few percent of the Eddington luminosity between a faint state characterized  by a truncated disc and a hot corona and  a bright state requiring  a warm corona in addition to a disc and a hot corona. In the case of IC~4329A, the disc truncation, the weakness of the soft excess component that contributes only $\sim 20\%$ to the broadband X-ray emission, and the low Eddington fraction $\sim 2.5\%$ for the  Czerny et al. extinction law strongly suggest that this AGN is very near the spectral transition point described by \cite{2018MNRAS.480.3898N} for the changing-look AGN. 

\astrosat{}/UVIT observations of AGN with negligible or known internal reddening will provide better constraints on the inner extent of the accretion disks.
We  plan to study the far and near UV \astrosat{} data on a number of bright AGN, which
 will allow us to disentangle the {\it intrinsic} AGN emission from the host galaxy flux in
 the near and far-UV bands, and investigate the inner accretion flow in these
 objects.

\section*{Acknowledgements}
We thank an anonymous referee for the constructive comments that improved the paper. This publication uses the data from the \astrosat{} mission of the Indian Space Research Organisation (ISRO), archived at the Indian Space Science Data Centre (ISSDC). This publication uses UVIT data processed by the payload operations centre at IIA. The UVIT is built in collaboration between IIA, IUCAA, TIFR, ISRO and CSA.  This research has made use of the SIMBAD database, operated at CDS, Strasbourg, France.   This research has made use of software provided by the Chandra X-ray Center (CXC) in the application package Sherpa.  PT acknowledges the University Grant Commission (UGC), Government of India for financial supports.

\section*{Data Availability}
The Level-1 UVIT data from the five observations with IDs 
9000001006, 9000001048, 9000001118, 9000001286 and 9000001340
used in this paper are publicly available at the \astrosat{} data archive \url{https://astrobrowse.issdc.gov.in/astro_archive/archive/Home.jsp} maintained by the ISSDC.


\bsp	
\label{lastpage}
\end{document}